\documentclass[prc,twocolumn,showpacs]{revtex4-1}
\pdfoutput=1 
\usepackage{setspace}
\usepackage{graphicx}
\usepackage{amsmath, amsthm, amssymb}
\usepackage{multirow}
\usepackage{subfigure}
\usepackage{float}
\begin{document}

\title{Precision spectral manipulation: a demonstration using a coherent optical memory}
\author{B. M. Sparkes, C. Cairns, M. Hosseini, D. Higginbottom, G. Campbell, P. K. Lam, and B. C. Buchler}
\affiliation{ Centre for Quantum Computation and Communication Technology, Department of Quantum Science, Research School of Physics and Engineering, The Australian National University, Canberra, ACT 0200, Australia}
\begin{abstract}
The ability to coherently spectrally manipulate quantum information has the potential to improve qubit rates across quantum channels and find applications in optical quantum computing. In this paper we present experiments that use a multi-element solenoid combined with the three-level gradient echo memory scheme to perform precision spectral manipulation of optical pulses. These operations include bandwidth and frequency manipulation, spectral filtering of separate frequency components, as well as time-delayed interference between pulses with both the same, and different, frequencies. These operations have potential uses in quantum information applications.
\end{abstract}

\maketitle

\section{Introduction}
\label{sec:ff_introduction}

Quantum information processing seeks to harness quantum mechanics to enhance information processing capabilities. Just as classical communication and computation requires memory buffers, quantum information systems will require memories for quantum states. An optical quantum memory allows coherent, noiseless and efficient storage and recall of optical quantum states. They are an essential building block for quantum repeaters \cite{bsduan}, which will extend the range of quantum communication. They could also find applications as a synchronization tool for optical quantum computers, and in a deterministic single-photon sources \cite{bslvovsky}. Much progress has been achieved towards this goal in recent years, with efficiencies up to 87\% \cite{bsnatcomm}, storage times of over one second \cite{bslongdell,bszhang}, as well as bandwidths above a gigahertz \cite{bsreim,bssaglamyurek} and over 1000 pulses stored at once \cite{bsbonarota}, all being separately demonstrated using different storage techniques.\\
If, however, we move towards manipulation of the stored information, a new range of possible uses for quantum memories appear. For instance, the ability to coherently manipulate the spectrum of pulses would prove a key tool for allowing quantum information transfer between systems with different bandwidths. This ability could also lead to increased bit rates over quantum communication channels \cite{bsmoiseev}. Another way of improving bit rates is the idea of multiplexing in quantum memories - a powerful tool where multiple signals are bundled into one over a communication channel. Multiplexing could be achieved with, for instance, different spatial \cite{bsvasilyev}, temporal \cite{bssimon} or frequency modes in a quantum memory. Being able to alter a pulse's shape, as well as its bandwidth, could also lead to increased bit rates due to a decrease in losses caused by pulse aberrations through various media (i.e. optical fiber \cite{bsfiber}).\\
Various coherent pulse manipulation techniques have already been demonstrated without the aid of a quantum memory. For instance, three-wave mixing \cite{bskielpinski}, quantum pulse gates \cite{bsbrecht,bseckstein}, and pulsed frequency up-conversion \cite{bsrikher} have all been shown to be able to coherently alter the temporal, and in some cases spectral, profile of  optical pulses. Work has also been carried out with  pulse shaping and splitting inside a coherent memory using electromagnetically-induced transparency (EIT) \cite{bsnovikova}.\\
In this paper we will investigate the coherent spectral manipulation abilities of the gradient echo memory (GEM) scheme \cite{bsprl108,bsalexander07}. GEM has been shown to have high efficiencies \cite{bsnatcomm, bshedges} and not add noise to the quantum state \cite{bsnatphys}, while also being able to store up to 20 pulses simultaneously \cite{bsnatcomm}, making it a promising candidate as an optical quantum memory. \\
Previous experimental work has shown that GEM is capable of manipulating stored light in a number of ways.  $\Lambda$-GEM, based on three-level atoms \cite{bsol}, has been used to resequence pulses, stretch or compress the bandwidth of stored pulses \cite{bsnature}, add a frequency offset to the recalled light \cite{bsnatcomm}, and interfere two pulses within the memory \cite{bsgeoff}. Modelling has shown that GEM is capable of much more. For example, it could be used as an optical router \cite{bscarreno2} or all-optical multiplexer \cite{bscarreno1}.
In this paper we investigate proposals that make particular use of the frequency encoding nature of GEM to coherently manipulate the spectrum of stored pulses, filter modulated pulses and combine or interfere pulses of different frequencies \cite{bsbb}.\\

The remainder of this paper is structured as follows: Sec. \ref{sec:ff_overview} presents an overview of the GEM protocol and relevant theory, before describing the experimental details in Sec. \ref{sec:ff_setup}. We then, in Sec. \ref{sec:ff_experiments} present experimental results characterizing basic frequency manipulation operations, as well as demonstrating frequency domain engineering with fine control of magnetic field gradients provided by a multi-element coil.\\
%Need to find good reference for fibers -> start from same point as kielpinski
\section{Gradient Echo Memory Overview}
\label{sec:ff_overview}

\begin{figure}[h]
\begin{center}
\includegraphics[width=\columnwidth]
{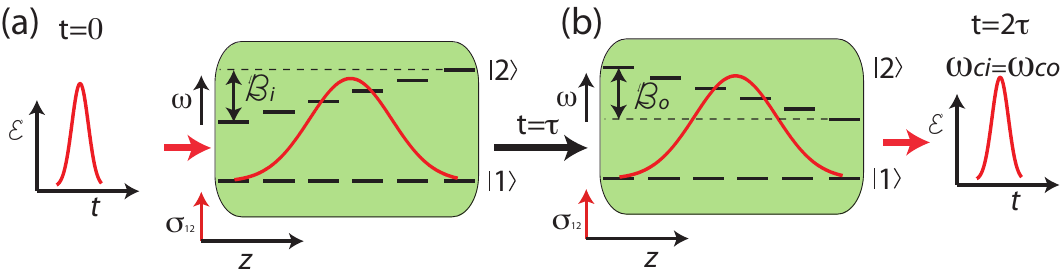} 
%{figure1.eps}
\caption{\textbf{Basic Gradient Echo Memory Operation}. (a) A pulse with an input Gaussian envelope $\mathcal{E}_i(t)$, bandwidth $\mathcal{B}_{p}$, and centre frequency $\omega_{ci}$ enters the storage medium at time $t=0$ where the optical information is stored in the atomic excitation $\sigma_{12}$. The memory has a linear frequency gradient placed along it in the $z$-direction and a input frequency bandwidth $\mathcal{B}_i$. (b) At time $t=\tau$ the sign of the frequency gradient is reversed, with the memory output bandwidth $\mathcal{B}_o=\mathcal{B}_i$. In this scheme the echo is emitted at time $t=2\tau$ with pulse shape $\mathcal{E}_o(t)=\mathcal{E}_i(-t)$ and centre frequency $\omega_{co}=\omega_{ci}$.}
\label{fig:ff_singlegrad}
\end{center}
\end{figure}

A linearly varying frequency gradient placed along an ensemble of atoms is the key component to the gradient echo memory scheme. The detuning of each atom from its original resonance, and therefore the frequency it will absorb, is proportional to its position along the memory. Therefore GEM is a frequency encoding memory, with pulses being stored as their spatial Fourier transform along the memory. For a linear gradient, $\eta(z) = \eta$, the bandwidth of the memory is determined by $\mathcal{B}_s = \eta l$, where $l$ is the length of the memory. We assume here that the centre pulse frequency $\omega_c$ is stored in the middle of the memory.\\
The basic two-level GEM operation is shown in Fig. \ref{fig:ff_singlegrad}. The equations that govern the storage of a light pulse with a slowly varying envelope operator $\hat{\mathcal{E}}(z,t)$ inside a two-level ensemble with atomic polarization operator $\hat{\sigma}_{12} \equiv \left|1\right\rangle\left\langle 2 \right|$ in this situation are \cite{bsprl108}
\begin{eqnarray}
\partial_t\hat{\sigma}_{12}& = &-\left[ \gamma + i\eta(z)\right]\hat{\sigma}_{12} + ig \hat{\mathcal{E}} \nonumber \\
\partial_z \hat{\mathcal{E}} & = & i\frac{gN}{c}\hat{\sigma}_{12},
\label{eq:ff_gemeqn}
\end{eqnarray}
where $\gamma$ is the decay rate from the excited state $\left|2\right\rangle$, $g$ is the coupling strength between the two levels, $N$ is the number of atoms, and $c$ is the speed of light. This equation assumes a weak probe field such that $\left\langle \sigma_{11} \right\rangle \approx 1$ holds, and that all atoms are initially in this ground state.\\
To recall the pulse, under normal GEM operation, the linear gradient is exactly reversed a time $\tau$ after the pulse has entered the memory, i.e. $\eta(t < \tau) = -\eta(t > \tau)$. This leads to a time-reversal of the absorption process described in Eq. \ref{eq:ff_gemeqn} and an emission of a time-reversed copy of the input pulse in the forwards direction, i.e. $\hat{\mathcal{E}}_o(t) = \hat{\mathcal{E}}_i(2\tau-t)$, at time $t=2\tau$, with the centre frequency of the echo being the same as the input pulse.\\
It is not necessary, however, to recall with an exact reversal of the input gradient to recall a pulse, or to have a constant gradient along the entire length of the memory. Indeed, having fine control of the input and output gradients, as discussed in the following sections, is what provides us with the ability to perform spectral manipulation operations using GEM.

\section{Experimental Set-Up} 
\label{sec:ff_setup}

\begin{figure}[h]
\begin{center}
\includegraphics[width=\columnwidth]
{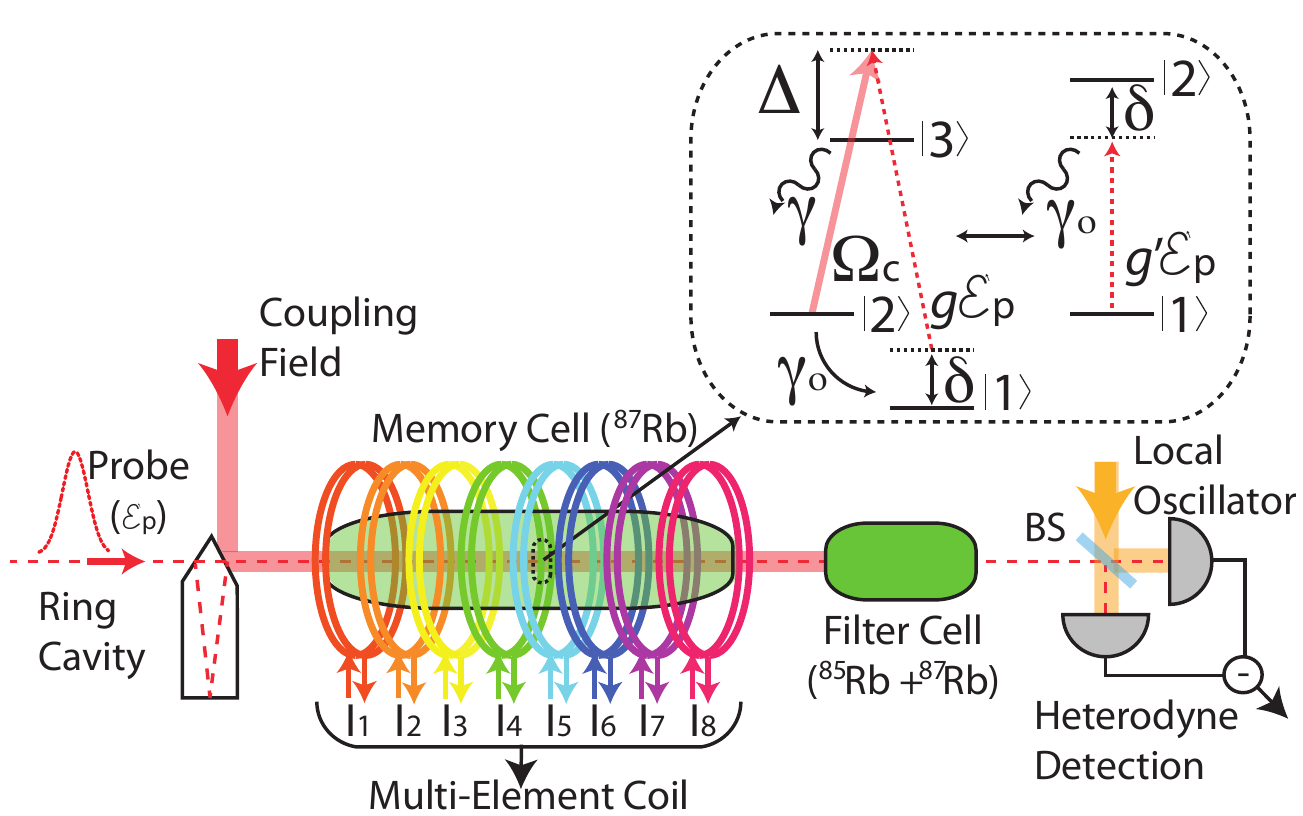} 
%{figure1.eps}
\caption{\textbf{Experimental set-up}. $\mathcal{E}_p$ - probe electric field envelope; BS - non-polarizing beam-splitter; and $I_{1}$-$I_{8}$ - currents supplied to the individual solenoids. Inset shows the level scheme and the equivalence between the $\Lambda$ system used and a two-level atom: $\Delta$ - one-photon detuning; $\delta$ - two-photon detuning; $\Omega_c$ - coupling field Rabi frequency; $\gamma$ - decay rate from the excited state; $\gamma_o$ - decoherence rate between the ground states; $g$ - coupling strength between ground and excited states; and $g'$ - effective coupling strength for the equivalent two-level system.}
\label{fig:ff_setup}
\end{center}
\end{figure}

An overview of the set-up used for the following spectral manipulation experiments is shown in Fig. \ref{fig:ff_setup}. We use the $\Lambda$-GEM scheme \cite{bsol} and warm rubidium-87 vapor for our experiments. This three-level system, where a strong classical field is used to couple the two ground states, is equivalent to the two-level one described in the previous section as long as (i) the one-photon detuning $\left|\Delta\right| \gg d \gamma$, and (ii) $1 \ll d T \gamma$ \cite{bsgorshkov}, where $d$ is the on-resonance optical depth of the system, $\gamma$ is the excited state decay rate, and $T$ is the fastest timescale of the system. This equivalence is shown in Fig. \ref{fig:ff_setup} inset. The coupling strength of the equivalent two-level system is given by $g' = g \Omega_c/\Delta$, where $\Omega_c$ is the Rabi frequency of the coupling field. The advantage of the $\Lambda$ system is that the storage time of the memory is now controlled by the ground state decoherence rate $\gamma_o$ which is much less than the excited state lifetime. Indeed there are a wide range of atoms with stable ground state configurations that are suitable for $\Lambda$-GEM. \\
The weak probe and strong coupling fields are derived from the same laser, which is blue detuned by approximately 3 GHz from the $F=2 \rightarrow F'=2$ $^{87}$Rb D1 transition. A small part of the laser is sent through a fibre-coupled electro-optic modulator driven at 6.8 GHz, the ground state splitting of $^{87}$Rb, and the positive sideband selected by passing it through a filtering cavity. This field, now 3 GHz blue detuned from the $F=1 \rightarrow F'=2$ transition, is used for both the probe and local oscillator (LO). The probe and control fields, having the same circular polarisation, are combined on a ring cavity that is resonant with the probe. The probe and coupling fields then enter the memory - a 25 mm diameter, 20 cm long gas cell containing isotopically enhanced $^{87}$Rb, and 0.5 Torr krypton buffer gas, heated to approximately 80$^{\circ}$C using an electronic filament heater. \\
Eight separate solenoid coils, with four turns each, are placed along the length of the memory. This multi-element coil (MEC) is used to create the complex gradients for the experiments discussed in the following sections by placing a different current in each coil, and using the superposition principle for magnetic fields, i.e. $B_{tot}(z)=\sum_{i}B_i(z)$. The two-photon detuning of each atom can then be defined as a function of position along the memory $\delta(z) = 2g_F \cdot B_{tot}(z) - \delta_{o}$, where $g_F = 0.7$ MHz/G is the Land\'{e} factor, and $\delta_{o}$ is an arbitrary two-photon offset (for instance, in this case $\delta_{o}=0$ is defined for a set dc magnetic field and coupling field frequency). The memory cell and coils are surrounded with a layer of $\mu$-metal to shield against external magnetic fields. The heater is turned off during the storage process to ensure there are no stray magnetic fields interacting with the atoms.\\
Upon leaving the memory, the probe and coupling fields pass through a filter cell containing a natural mixture of Rb (i.e. $^{85}$Rb and $^{87}$Rb) and  heated to approximately 150$^{\circ}$C. Due to the detunings chosen above, the coupling field is resonant with a $^{85}$Rb transition, leading to approximately 40 dB suppression through the cell. The probe field, which passes through the filter cell with 70\% efficiency, is then combined with the local oscillator signal on a non-polarizing beam-splitter and heterodyne detection is performed. Fine control of the frequencies of all fields, as well as gating of the probe and coupling fields, is achieved using acousto-optic modulators. This experiment is controlled with an augmented version of the LabVIEW$\textregistered$ code presented in Ref. \cite{bsrsi}.

\section{Spectral Manipulation Experiments}
\label{sec:ff_experiments}
In this section, we present the various spectral manipulation experiments undertaken with the set-up presented in the previous section.

\subsection{Centre Frequency Manipulation}
\label{sec:ff_frequency}

\begin{figure}[h]
\begin{center}
\includegraphics[width=\columnwidth]
{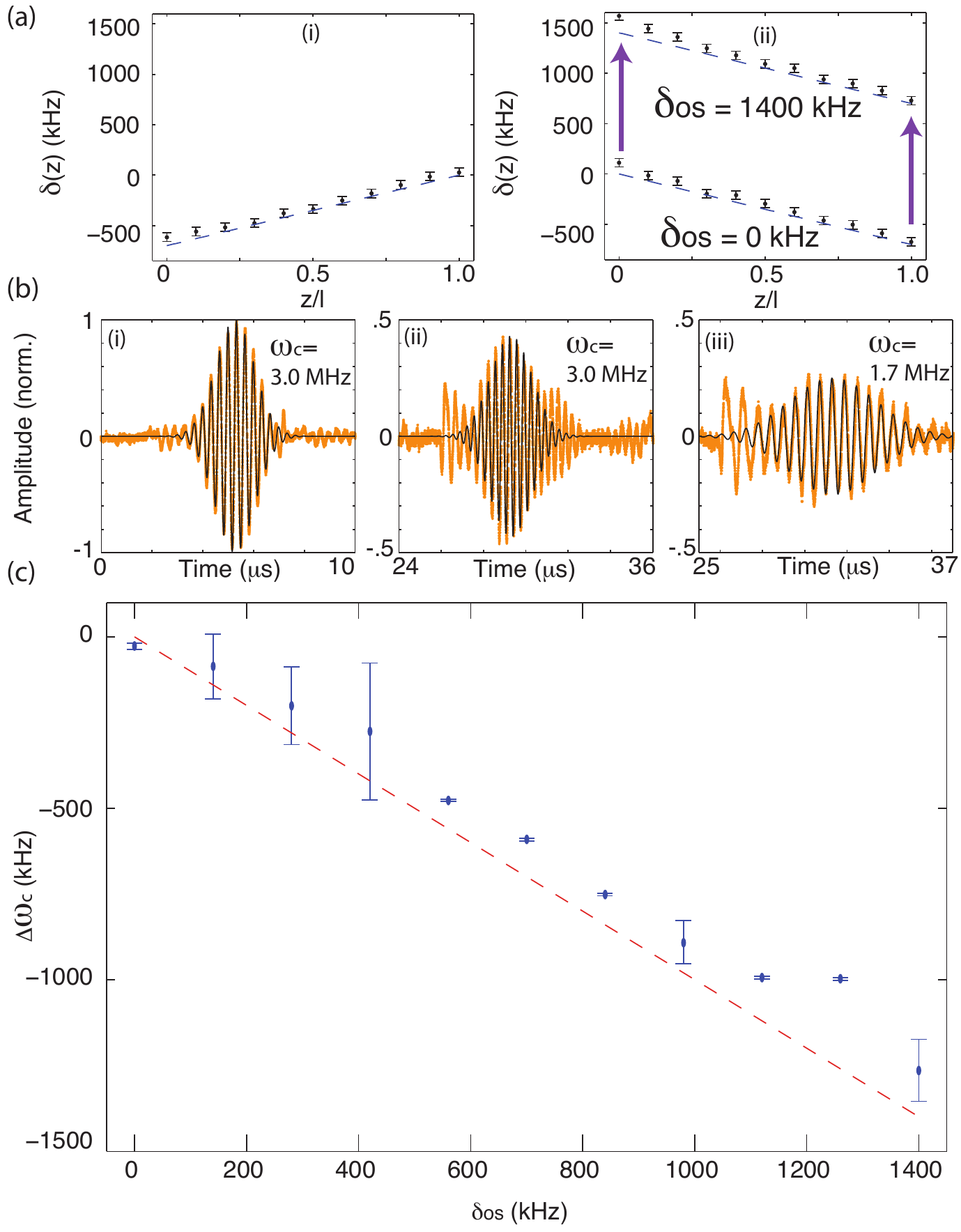} 
\caption{\textbf{Manipulating pulse frequency}. (a) Two-photon detuning $\delta$ as a function of position $z$ along the memory (normalized to length $l$) due to (i) input gradient and (ii) output gradients with minimum and maximum gradient offset $\delta_{os}$ (as noted on figure). Blue (dashed) line corresponds to the desired field and points correspond to the measured magnetic field (error bars due to sensitivity of Gauss-meter). (b) Heterodyne data showing (i) input pulse; (ii) echo for recall with $\delta_{os}=0$; and (iii) echo for recall with $\delta_{os}=1400$ kHz offset. Orange points correspond to raw data, black lines correspond to modulated Gaussian fit to data, and $\omega_c$ values correspond to the centre frequencies of pulses extract from the fits. (c) The change in centre frequency of the output pulses relative to the input pulse $\Delta\omega_c$  as a function of $\delta_{os}$. Points represent the measured centre frequency (error bars from standard deviation of 100 traces), and the dashed line corresponds to the theoretical behaviour.}
\label{fig:ff_gradoffset}
\end{center}
\end{figure}

By adding an offset to the recall field, i.e. $\delta(z,t > \tau) = -\delta(z,t < \tau) + \delta_{os}$, the centre frequency of the echo relative to the input can be altered and will be given by
\begin{equation}
\omega_{co} = \omega_{ci} \pm \delta_{os}.
\label{eq:ff_gradoffset}
\end{equation}
The sign of $\delta_{os}$ is dependent on which ground state is used. In this case it is the $\left|F=1,m_F=1\right\rangle$ state and therefore the sign will be negative. Figure \ref{fig:ff_gradoffset}(a) shows plots of (i) input and (ii) output gradients with varying magnetic field offsets. Figure \ref{fig:ff_gradoffset}(b) shows single heterodyne traces for (i) input and (ii) echo with no applied offset, as well as (iii) echo with an applied offset of 1.4 MHz.\\
The stretched form of the echoes indicates that there is dispersion present in the memory. This is not surprising considering that the bandwidth of the memory is only slightly greater than the bandwidth of the pulse. This effect is accentuated for longer storage times. The additional elongation for recall with a greater offset indicates that fringes of the magnetic field (i.e. those components that tail off at either end of the cell) may have affected the stored pulse, leading to greater dispersion. We note, however, that such effects are easily compensated, as we explain in Section \ref{sec:ff_bandwidth}.\\
A modulated Gaussian was fitted to the main body of the output pulses in order to extract the value of $\omega_c$ relative to the LO frequency. This is also shown in Fig. \ref{fig:ff_gradoffset}(b) for the input, as well as the two echoes. Figure \ref{fig:ff_gradoffset}(c) shows a characterization of the change in $\omega_c$ for a range of values of $\delta_{os}$. This is compared with the behaviour expected from Eq. \ref{eq:ff_gradoffset}. As can be seen, the two are in good agreement.\\

\subsection{Bandwidth Manipulation}
\label{sec:ff_bandwidth}

\begin{figure}[h]
\begin{center}
\includegraphics[width=\columnwidth]
{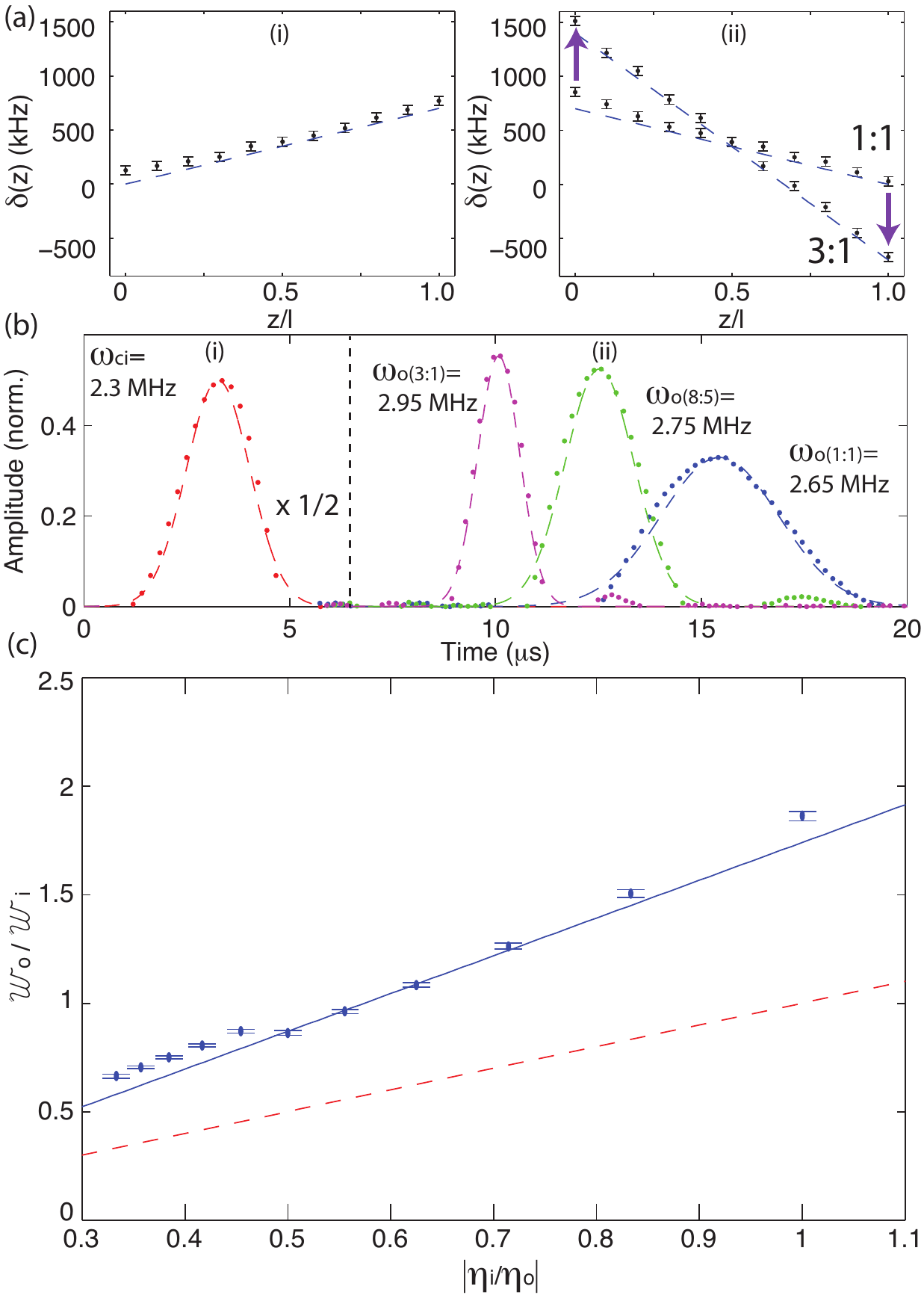} 
\caption{\textbf{Manipulating pulse bandwidth}. (a) Two-photon detuning $\delta$ as a function of position $z$ along the memory (normalized to length $l$) due to (i) input gradient and (ii) minimum and maximum output gradients (ratios noted on figure). Blue (dashed) line corresponds to the desired field and points correspond to the measured magnetic field (error bars due to sensitivity of Gauss-meter). (b) Amplitude plot, normalised to size of input pulse, showing (i) input pulse (shown in red, scaled by a factor of $1/2$), (ii) output pulses recalled with varying output gradients as noted. Points correspond to demodulated data, dashed lines correspond to Gaussian fit to data, and $\omega_{c}$ values correspond to the centre frequencies of pulses relative to the LO. Bracketed ratios indicate $\eta_o/\eta_i$. (c) The FWHM of the output pulses $\mathcal{W}_o$ normalized to the FWHM of the input pulse $\mathcal{W}_i$, as a function of input gradient over output gradient $\left|\eta_i/\eta_o\right|$. Points represent measured FWHM (error bars from standard deviation of 100 traces), red (dashed) line corresponds to Eq. \ref{eq:ff_grad}, blue (solid) line corresponds to linear fit to data.}
\label{fig:ff_gradchange}
\end{center}
\end{figure}

By recalling with a steeper output gradient than input gradient, i.e. $\left|\eta(z, t > \tau) \right| > \left| \eta(z, t < \tau) \right|$, the output bandwidth of the memory $\mathcal{B}_o$ will be made greater than the input bandwidth. This change in bandwidth is, in turn, passed on to the echo as the absolute emission frequency relative to the centre frequency $\left| \omega(z) - \omega_c\right|$  of each atom along the ensemble will be greater, while the total excitation $\sigma_{12}(z,t)$ will remain unchanged. In this case, the output pulse will be compressed in time due to its now greater frequency spectrum. The opposite is also true, i.e. by recalling with a shallower gradient the output pulse bandwidth will be reduced and the pulse elongated in time. \\
The temporal profile of the output pulse, measured using the pulse full-width-half-maximum (FWHM) $\mathcal{W}_o$ can be simply expressed as a function of the input profile $\mathcal{W}_i$ and input/output gradient $\eta_{i/o}$ as
\begin{equation}
\mathcal{W}_o = \mathcal{W}_i \cdot \left| \eta_i/\eta_o \right|.
\label{eq:ff_grad}
\end{equation}
\\
This has already been experimentally demonstrated \cite{bsnature}. Here, however, we present a more quantitative study with the extra control of the gradient we obtain with the MEC. Figure \ref{fig:ff_gradchange}(a) shows experimental plots of (i) the input gradient, and (ii) output gradients with ratios $\left|\eta_o/\eta_i\right|$ from 1:1 to 3:1. Performing fits to individual pulses, as illustrated in Fig. \ref{fig:ff_gradoffset}(b), allows for in-phase digital demodulation of the heterodyne data. This, in turn, allows for averaging over many traces, something that would not be possible with the non-demodulated data due to phase fluctuations between the probe and local oscillator.\\
Figure \ref{fig:ff_gradchange}(b) shows averaged demodulated input and output pulse amplitudes for different recall gradients. As predicted, the output pulses become more compressed as the recall gradient is increased. Though it does follow a linear relationship, it does not, however, follow Eq. \ref{eq:ff_grad}, as can be seen from Fig. \ref{fig:ff_gradchange}(c). This discrepancy is most probably a result of the highly dispersive nature of GEM storage, with large changes in the absorptive profile of the system especially at either side of the GEM frequency storage window when the pulse bandwidth is approximately equal to the memory bandwidth, as discussed in the previous section. Having a larger input bandwidth would reduce the effect of dispersion on the pulse.\\
It can also be seen from Fig. \ref{fig:ff_gradchange}(b) that the echoes are emitted from the memory earlier, i.e. at a time $t < 2\tau$, when recalled with a steeper output gradient. This is because a steeper gradient will cause the rephasing process to occur at a faster rate, and will also affect the amount of dispersion. The frequency of the echo is not the same as the input pulse due to the inherent GEM frequency shift predicted in Ref. \cite{bsmoiseev2}, which is greater for shorter storage times.\\ 

\subsection{Spectral Filtering}
\label{sec:ff_filtering}

\begin{figure}[h]
\begin{center}
\includegraphics[width=\columnwidth]
{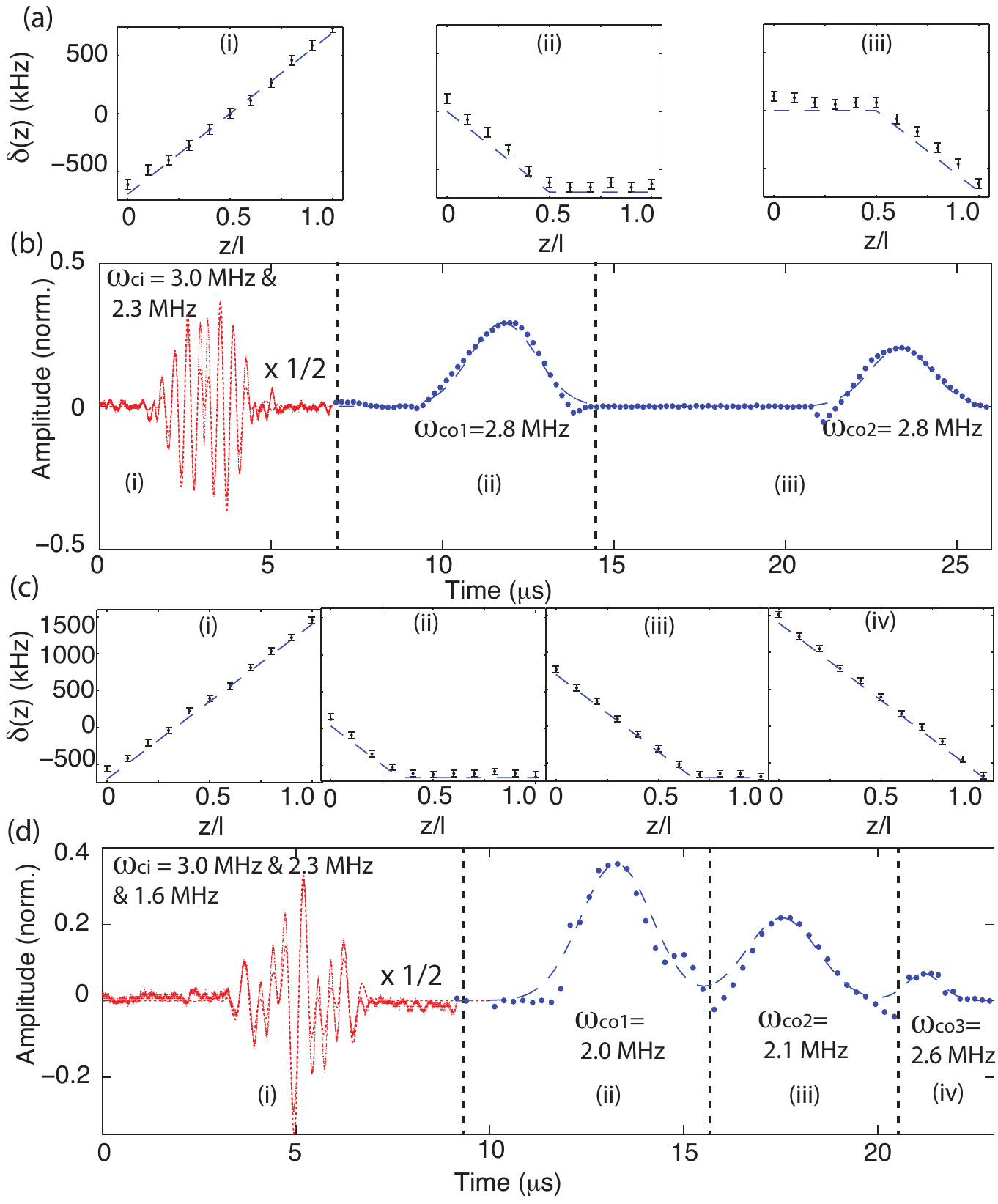} 
\caption{\textbf{Spectral filtering}. (a) Two-photon detuning $\delta$ as a function of position $z$ along the memory (normalized to length $l$) due to gradients (i)-(iii) corresponding to times (i)-(iii) in (b). For traces (a) and (c) blue (dashed) lines correspond to the desired field and points correspond to the measured magnetic field (error bars due to sensitivity of Gauss-meter). (b) Spectral filtering of (i) a Gaussian envelope containing two frequency components separated by 700 kHz (red, non-demodulated, scaled by $1/2$), and the demodulated retrieval (blue) of (ii) higher,  and (iii) lower frequency components averaged over 100 traces. For traces (b) and (d) points correspond to data, lines correspond to fit to data, and $\omega_c$ values correspond to centre frequencies of pulses. (c) Two-photon detuning due to (i) input and (ii), (iii), and (iv) output gradients corresponding to times (i)-(iv) in (d), which shows the conversion from the time to frequency domain of (i) a Gaussian pulse with two modulation sidebands at $\pm 700$ kHz (red, non-demodulated, scaled by $1/2$), and the demodulated retrieval of (ii) higher frequency sideband, (iii) carrier, and (iv) lower frequency sideband averaged over 100 traces (blue).}
\label{fig:ff_spectralfiltering}
\end{center}
\end{figure}

If we now consider the storage of a modulated pulse, the frequency encoding nature of GEM will mean that the carrier and sideband components of the pulse will be stored in different parts of the memory. Therefore, if we had fine enough control over the recall gradient, we could choose when to recall the different frequency components by switching the gradient only in the pertinent part of the memory. \\
An experimental demonstration of this filtering is shown in Fig.s \ref{fig:ff_spectralfiltering}(a)-(b). Here a carrier pulse with a Gaussian envelope and two frequency components separated by 700 kHz is sent into the memory. By reversing the gradient only in one half of the memory at a time, the different frequency components of the pulse can be recalled separately. The output pulses in this case both have the same $\omega_{co}$ due to the offset in the recall gradient for the lower frequency component.\\
Furthermore, following the same logic, by being able to switch the gradient slowly along the length of the memory, the stored pulse can be recalled as its Fourier transform. This is shown experimentally in Fig.s \ref{fig:ff_spectralfiltering}(c)-(d) for a modulated Gaussian with two sidebands at $\pm 700$~kHz. In this case the gradient is reversed in three stages, rather than a gradual reversal of the entire memory, due to the limitations of the LabVIEW$\textregistered$ code refresh rate ($\approx1$ $\mu$s). The three outputs were fit separately to allow for their different frequencies. The time window for each demodulation is denoted by the dashed lines in Fig. \ref{fig:ff_spectralfiltering}(d).\\
In the above experimental demonstration, equal power was put into the sidebands and carrier. The reason this is not the case for the echo is because of coupling field-induced scattering of light. In normal $\Lambda$-GEM storage the coupling field is switched  off during the storage process to limit this effect. This is not possible, however, for multi-pulse recall in a single gas cell as the coupling field must be present for recall to occur.\\

\subsection{Pulse Interference}
\label{sec:ff_interference}

\begin{figure}[h]
\begin{center}
\includegraphics[width=\columnwidth]
{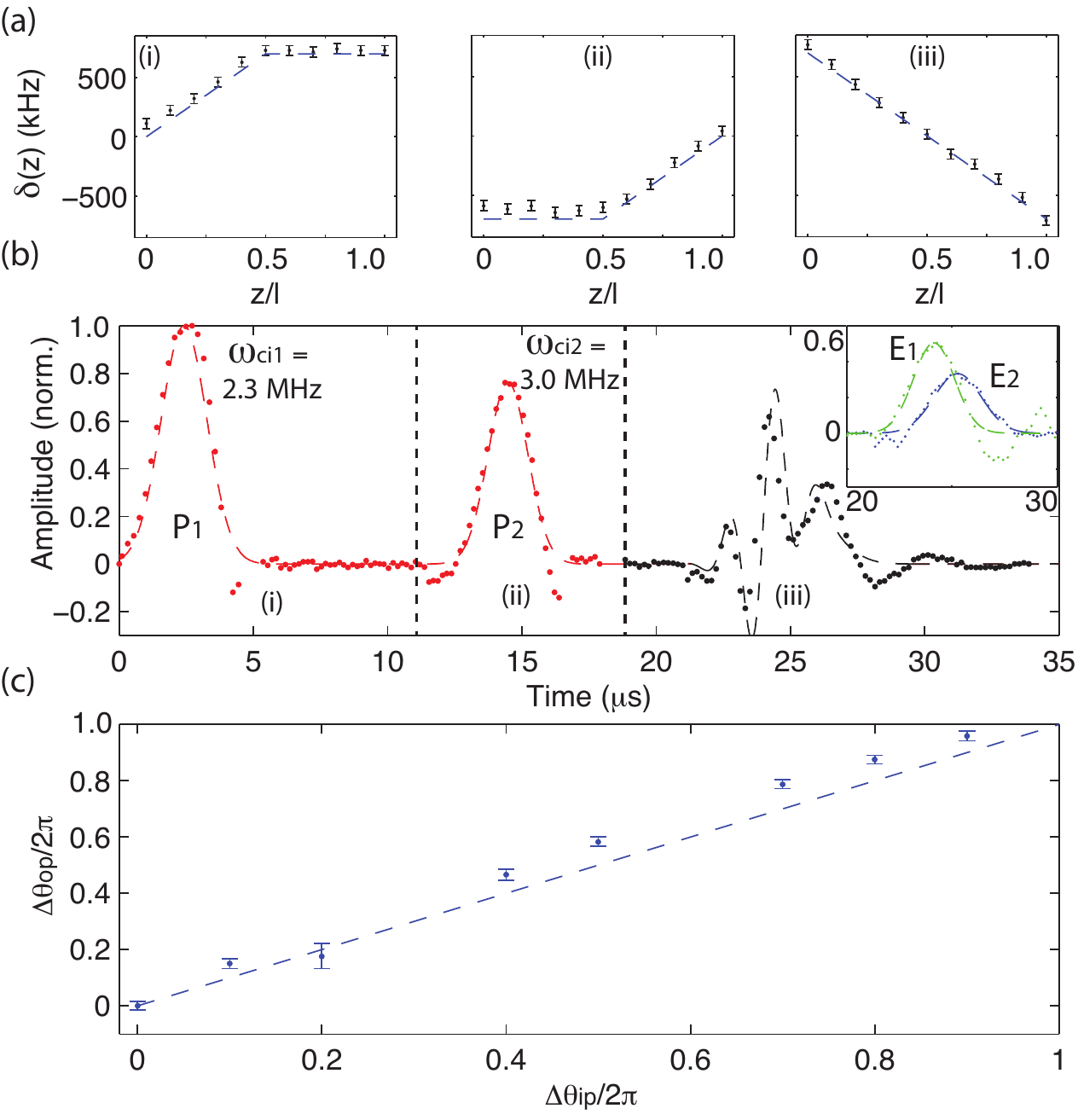} 
\caption{\textbf{Interference with pulses of different frequencies}. (a) Two-photon detuning $\delta$ as a function of position $z$ along the memory (normalized to length $l$) due to (i)-(ii) input gradients and (iii) output gradient corresponding to times (i)-(iii) in (b). Blue (dashed) lines correspond to the desired field and points correspond to the measured magnetic field (error bars due to sensitivity of Gauss-meter). (b) Interference of two initially time separated pulses (i) $P_1$ and (ii) $P_2$, shown in red, which are also separated in frequency by 700 kHz. (iii) Shows the superposition of the two pulses. The inset shows the output from the memory for storage of only a single pulse: $P_1$ recall ($E_1$, green); or $P_2$ recall ($E_2$, blue). Points correspond to demodulated data averaged over 100 traces, lines correspond to Gaussian fit to data, and $\omega_c$ values correspond to centre frequencies of pulses. (c) The change in relative phase of the fitted interference pulse $\Delta\theta_{op}$ as a function of the relative phase of the input pulses $\Delta\theta_{ip}$. Points represent data extracted from fit (error bars from standard deviation of 100 traces), and the dashed line corresponds to the theoretical behaviour.}
\label{fig:ff_difffreqint}
\end{center}
\end{figure}

A time reversal of the spectral filtering process is also possible. That is, if we take two pulses with different frequencies and store them one at a time in different halves of the memory we can alter the gradients in the different halves at different times. This will cause the recalled echoes to overlap, and therefore interfere, at the output of the memory. Previous experiments in pulse interference using GEM have shown how the memory can facilitate interference between modes separated in either the time or frequency domains \cite{bsgeoff}. Here we look at an alternate method using complex gradients made possible with the MEC.\\
This is shown in Fig.s \ref{fig:ff_difffreqint}(a)-(b). Here, two pulses separated in frequency by 700 kHz are stored in separate halves of the memory. The lower (higher) frequency pulse being stored in the second (first) half. Setting the gradient to 0 in the second half of the memory when the first (lower frequency) pulse $P_1$ enters ensures it will be stored in the first half of the memory. Setting the gradient to 0 in the first half of the memory while the higher frequency pulse $P_2$ enters and is stored serves two purposes: apart from ensuring that none of $P_2$ is stored in the first half of the memory, it also means that the stored $P_1$ will not undergo any additional dephasing. Therefore, after $P_2$ is stored we can reverse the gradient across the entire memory at once, causing the superposition of the echoes on the output.\\ 
To investigate the phase preserving quality of the memory, we altered the relative phase between $P_1$ and $P_2$ and looked at the phase of the interference pattern of the echo. As can be seen from Fig. \ref{fig:ff_difffreqint}(c), the change in relative phase of the two input pulses matches the relative phase of the interference pattern at the output. The only free parameter in the fitting of the echoes was the relative phase, with the amplitude, timings and frequencies of the two individual pulses taken from storage of individual echoes $E_1$ and $E_2$, shown in Fig. \ref{fig:ff_difffreqint}(b) inset.\\

\begin{figure}[h]
\begin{center}
\includegraphics[width=\columnwidth]
{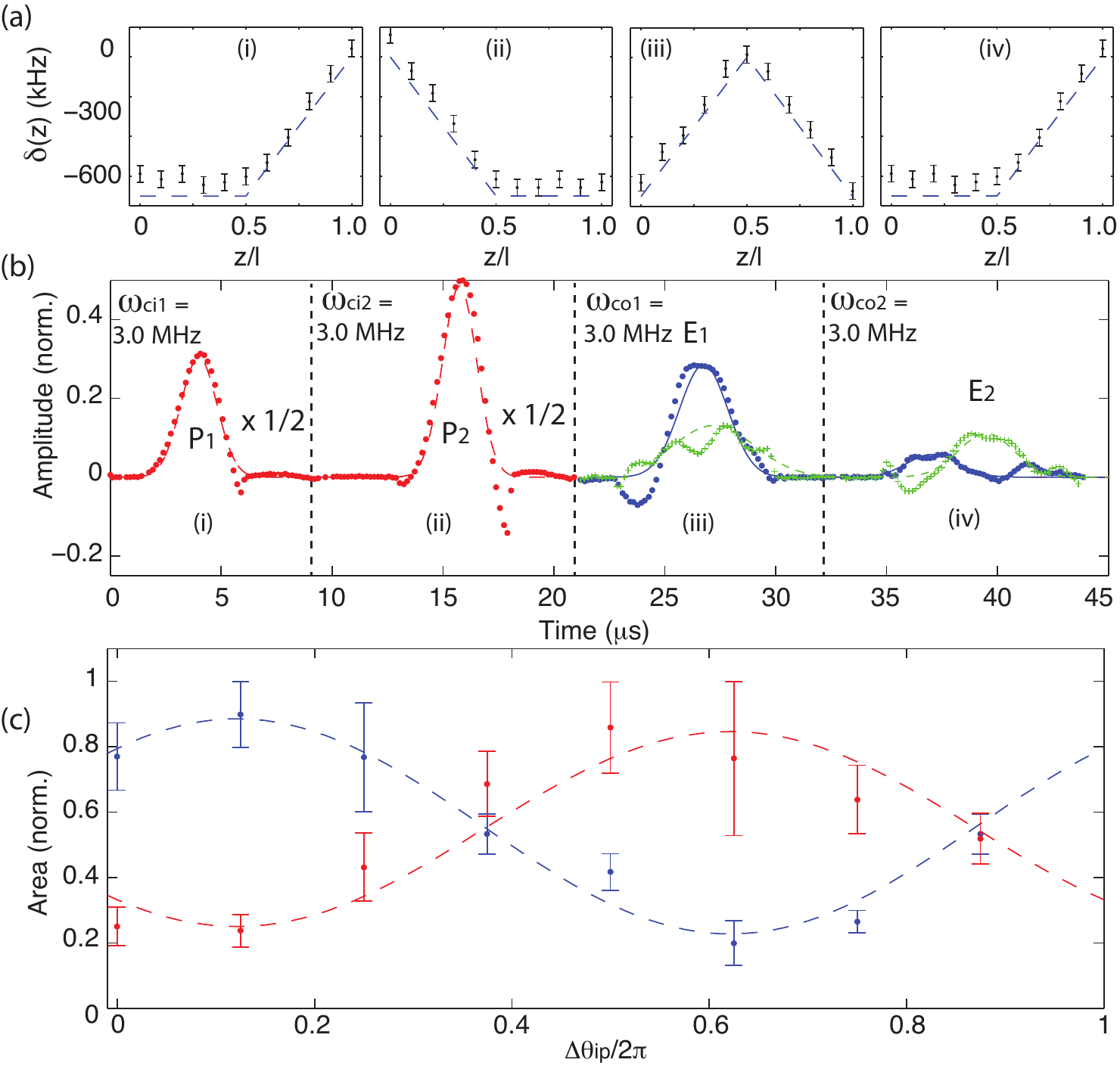} 
\caption{\textbf{Interference with pulses of the same frequency}. (a) Two-photon detuning $\delta$ as a function of position $z$ along the memory (normalized to length $l$) due to (i)-(ii) input gradients and (iii)-(iv) output gradients corresponding to times (i)-(iv) in (b). Blue (dashed) lines correspond to the desired field and points correspond to the measured magnetic field (error bars due to sensitivity of Gauss-meter). (b) Interference of two initially time separated pulses (i) $P_1$ and (ii) $P_2$ (red dashed lines), which have the same centre frequency. (iii) Initial $E_1$, and (iv) secondary $E_2$ superpositions of the two pulses. Blue points and solid line correspond to maximum constructive interference for $E_1$, while green crosses and dashed line correspond to maximum destructive interference for $E_2$. Points correspond to demodulated data averaged over 100 traces, lines correspond to Gaussian fit to data, and $\omega_c$ values correspond to centre frequencies of pulses. (c) The change in area (normalized to the maximum intensity of the individual echoes) as a function of the relative phase of the input pulses $\Delta\theta_{ip}$ for (i) $E_1$, and (ii) $E_2$. Points represent data extracted from fit (error bars from standard deviation of 100 fits), and the dashed line corresponds to a fit to the data.}
\label{fig:ff_samefreqint}
\end{center}
\end{figure}

Two pulses with the same frequency can also be interfered in a similar manner. An experimental demonstration of same frequency interference is shown in Fig. \ref{fig:ff_samefreqint}. The two pulses are, as before, stored in different halves of the memory by setting $\eta=0$ in the other half. As can be seen from Fig. \ref{fig:ff_samefreqint}(a)(iii), the first recall gradient  is no longer monotonic and, therefore, the pulse stored in the first half of the memory $P_2$ will be partly re-absorbed in the second half. This is why $P_2$ has a greater amplitude than $P_1$. The non-absorbed component of $P_2$ will interfere with the retrieved light from the second part of the memory. If these two echoes are in phase then there will be constructive interference and an enhanced echo $E_1$ will be retrieved. If they are out of phase then $E_1$ will be small and the residual energy will remain as atomic excitation inside the memory. Therefore, if the gradient in the second half of the memory is switched again, a second echo $E_2$ can be recalled from the memory from these leftover excitations. This is shown in Fig.s \ref{fig:ff_samefreqint}(b)(iii)-(iv).\\
Figure \ref{fig:ff_samefreqint}(c) shows the areas of the two echoes as the relative phase between $P_1$ and $P_2$ is varied. Interference fringes can be seen, with a $\pi$ phase shift between the two echoes. The visibility for both $E_1$ and $E_2$ is approximately 60\% (normalized to maximum echo output for $E_1$ and $E_2$ separately). These values could potentially be improved with finer gradient control, especially in terms of timings.\\

\section{Discussion}
\label{sec:ff_discussion}
In this paper we have experimentally demonstrated a number of different spectral manipulation operations using GEM. These operations could have various uses in a quantum information network. For instance, the ability to alter the bandwidth of the pulse, demonstrated in Section \ref{sec:ff_bandwidth}, could be used to match systems with different bandwidths and, by increasing bandwidths, help to improve bit rates for a number of time-bin qubits as described in \cite{bsmoiseev}. Combining this with the ability to change the centre frequency of the stored information, demonstrated in Section \ref{sec:ff_frequency}, would allow one, in theory, to match any two optical systems.\\
This latter ability would allow for the conversion of time-bin qubits into frequency-bin qubits. It would also, along with the frequency encoding nature of GEM, allow for the ability of frequency multiplexing. A number of pulses with different frequencies could be combined into one temporal pulse inside the memory, as demonstrated in Section \ref{sec:ff_interference}, and sent down the communication channel. Once they reached the other end of the channel they could be separated with a second memory, as demonstrated in Section \ref{sec:ff_filtering}. This could greatly improve qubit rates over optical channels in quantum information networks. Multiplexing quantum memories and nodes has also been suggested as a way of improving quantum repeater designs by speeding up the entanglement generation process \cite{bssimon}, and against memory coherence times \cite{bscollins}.\\
The phase sensitive interference of initially time separated pulses, demonstrated in Section \ref{sec:ff_interference}, could also find applications in quantum computing. In an all optical switch, for instance, where it is the relative phase between the pulses that determines how much light is emitted at different times \cite{bsharris}.\\
All these potential applications require high efficiencies and therefore high optical depths. A high optical depth is especially important as increasing the bandwidth of the system will decrease the recall efficiency. Also, a drawback to using one physical memory (i.e. the 20 cm long gas cell) as a system of $n$ sub-memories is that the optical depth for each individual sub-memory will be $1/n$ of the total memory optical depth.\\
An alternative method would be to use $n$ physical memories placed in series to create a memory network. Not only would this increase the overall optical depth of the system but it could help to alleviate two other drawbacks to the sub-memory approach taken here: finer control of the gradient; and coupling field-induced scattering. Much care was taken with the construction of the multi-element coil, and the decision on the order of gradients, to ensure the desired and physical gradients matched as well as possible. This does, however, place limitations on the operations that can be applied. Using many physical memories as one memory network would automatically increase the resolution of the gradients with respect to the length of each sub-memory. Another option for improving the resolution would be to move to an alternate gradient creation technique such as the ac Stark effect \cite{bsacs}.\\
The coupling field-induced scattering was discussed in Section \ref{sec:ff_filtering} with regards to the extra decay of information that is left in the memory while other information is recalled. This is a concern as this scattering leads to a decoherence rate almost 10 times larger than other decoherence mechanisms present \cite{bsnatcomm} and cannot be combatted in a single memory. However, this issue could be addressed with a network of memories if one were to use orthogonal polarizations for the probe and coupling fields and place polarizing beam-splitters between the memories.\\

\section{conclusions}
In this paper we have presented experimental demonstrations of theoretical spectral manipulation operations originally investigated in Ref. \cite{bsbb}. We showed that using the gradient echo memory scheme we can alter the bandwidth (and therefore temporal profile) of a pulse, as well as change its centre frequency. We also demonstrated the ability of GEM to act as a spectral filter and, using the frequency-encoding nature of GEM, were able to recall a modulated pulse as its Fourier transform. Finally we showed that two initially time separated pulses, with the same or different frequencies, could be caused to interfere coherently at the output of the memory. These abilities could be used to improve qubit rates across quantum communication channels, as well as potential uses in quantum computing applications.\\

\section{Acknowledgements}
Many thanks to Shane Grieves for his work constructing the hardware for the MEC, and to Peter Uhe for their initial testing. This research was conducted by the \textit{Australian Research Council Centre of Excellence for Quantum Computation and Communication Technology} (project number CE110001027).

\end{document}